\documentstyle[aps,twocolumn]{revtex}

\begin{document}
\draft
\title{
Proposal for
teleportation of the wave function of a massive particle
}

\author{A.S. Parkins$^1$ and H.J. Kimble$^2$}
\address{
$^1$Department of Physics, University of Auckland, Auckland, 
New Zealand
}
\address{
$^2$Norman Bridge Laboratory of Physics 12-33, California Institute of
Technology, Pasadena, CA 91125, U.S.A.
}

\date{\today}

\maketitle
\begin{abstract}
We propose a scheme for teleporting an atomic center-of-mass wave
function between distant locations. The scheme uses interactions
in cavity quantum electrodynamics to facilitate a coupling between
the motion of an atom trapped inside a cavity and external 
propagating light fields. This enables the distribution of
quantum entanglement and the realization of the required 
motional Bell-state analysis.
\end{abstract}



In a landmark work of 1993, Bennett {\em et al}. 
\cite{Bennett93} discovered a procedure for teleporting
an unknown quantum state from one location to another. 
The essential ingredient in their protocol is quantum 
entanglement of a bipartite system shared by the sender, Alice, 
and receiver, Bob. This shared entanglement, in unison with suitable 
measurements performed by Alice and communicated via 
classical channels to Bob, `mediates' the state 
transfer. 

Since the work of Bennett {\em et al}., a variety of 
possible experimental schemes for the 
teleportation of quantum states of {\em two-state} systems
have been proposed, in large part by the quantum optics 
community (see, e.g., 
\cite{Davidovich94,Cirac94,Sleator95,Braunstein95}).
In an exciting recent development, the first experimental 
investigations of teleportation of such states have 
been performed \cite{Boschi98,Bouwmeester97} with, in 
particular, the polarization state of a photon providing
the two-state system of interest.

Complementing this work on two-state systems has been
research into the teleportation of states of 
infinite-dimensional systems 
\cite{Vaidman94,Braunstein98a}, culminating last year
in the experimental demonstration of quantum teleportation 
of {\em optical coherent states} \cite{Furusawa98}.
This experiment was based on the specific proposal of 
\cite{Braunstein98a}, utilizing 
squeezed-state entanglement and balanced homodyne 
measurements of the light fields. 
Given that the experiment employed only standard optical
elements and measurement techniques, 
it offers significant promise of further intriguing
possibilities for quantum information processing with
continuous quantum variables, including quantum dense
coding \cite{Braunstein99a}, and
universal quantum computation \cite{Lloyd99}.

Another burgeoning field of research in quantum information
science is the implementation of quantum logic 
with trapped atoms or ions. Inspired by the proposal of
Cirac and Zoller \cite{Cirac95a} for a quantum computer 
based on the motional and internal degrees of freedom of 
a collection of trapped ions, impressive experimental 
progress has been made towards controlling quantum 
properties in such systems 
\cite{Monroe95,Wineland98}.
Particular advantages of trapped atom systems 
include long coherence times and the exquisite control of
transformations between motional and internal states.

In view of this tremendous
potential of both light- and motion-based schemes
for quantum information processing, it is sensible to 
investigate possibilities for combining the two
approaches and their distinct advantages. 
One particular application would be to quantum 
networks for distributed quantum computing and communication,
where, e.g., the `distribution' is accomplished with light 
fields \cite{Cirac98}, while
local processing is performed on motional states of a 
collection of trapped atoms.
Indeed, 
with the protocols of \cite{Cirac98} in mind, 
we have recently proposed and analyzed a cavity-QED-based system 
that enables the transfer of quantum states between the 
motion of a trapped atom and propagating light fields 
\cite{Parkins99a,Parkins99b},
which should lead to new capabilities for the synthesis and 
control of quantum states for both motion and light.

A particular example from \cite{Parkins99b} is the possibility
of creating an EPR (Einstein-Podolsky-Rosen) state in 
position-momentum for distantly separated atoms. The creation
of such an entangled state between remote particles suggests an
avenue for achieving teleportation of an unknown wave function of
a trapped particle between the EPR sites. In this letter, we propose
and analyze one such protocol that enables the teleportation of
an unknown one-dimensional {\em atomic center-of-mass wave function}
and that should be attainable within the context of emerging
experimental capabilities for trapping atoms in cavity QED
\cite{Hood98,Munstermann99}.

Our proposed teleportation scheme is shown schematically in
Fig.~1. 
Each of Alice (${\cal A}$), Bob (${\cal B}$), and Victor (${\cal V}$) 
possess an atom trapped inside an optical cavity. The aim is to
teleport the ($x$-dimension) motional state of Victor's atom to Bob's 
atom. This is achieved via 
the three usual stages for continuous quantum variables 
\cite{Braunstein98a,Furusawa98} --
(i) preparation of quantum entanglement between Alice and Bob's
atoms, (ii) Bell-state (homodyne) measurement by Alice, and (iii) 
phase-space displacement [$D(\alpha^\ast )$] by Bob (given the 
classical result $\alpha$  
of Alice's measurement). Finally, the state of Bob's atom may be 
examined by Victor for verification of the quality of the teleportation
(or indeed, physically delivered to Victor).
Stages (i) and (ii), as illustrated in Fig.~1, employ the 
cavity-mediated motion-light state transfer scheme of 
\cite{Parkins99a}, to which we now turn our attention.

Briefly, a single two-level atom (or ion) is tightly confined
in a harmonic trap located inside a high-finesse optical 
cavity. The atomic transition of frequency
$\omega_{\rm a}$ is coupled to a single mode of the cavity field of
frequency $\omega_{\rm c}$ and also to an
external (classical) laser field of frequency $\omega_{\rm L}$
and strength ${\cal E}_{\rm L}$.
The physical setup and excitation scheme are depicted in Fig.~2(a). 
The cavity is aligned along the $x$-axis, while
the field ${\cal E}_{\rm L}$ is incident from a direction in the 
$y$-$z$ plane.
Both the cavity field and ${\cal E}_{\rm L}$ are far from resonance 
with the atomic transition, but their difference frequency is chosen
so that they drive Raman transitions between neighboring 
motional number states (i.e., $\omega_{\rm c}-\omega_{\rm L}=\nu_x$,
with $\nu_x$ the $x$-axis trap frequency).

A number of assumptions are made in order to achieve the desired 
motion-light coupling: 
(i) Atomic spontaneous emission is neglected and 
the internal atomic dynamics adiabatically eliminated.
(ii) The size of the harmonic trap, located at a {\em node} of the cavity
field, is taken to be small compared to the
optical wavelength (Lamb-Dicke regime), enabling the approximations
$\sin (k\hat{x})\simeq \eta_x (\hat{b}_x+\hat{b}_x^\dagger )$
and ${\cal E}_{\rm L}(\hat{y},\hat{z},t)\simeq
{\cal E}_{\rm L}(t){\rm e}^{-i\phi_{\rm L}}$, where
$\eta_x$ ($\ll 1$) is the Lamb-Dicke parameter
and $\hat{x}=(\hbar /2m\nu_x)^{1/2}(\hat{b}_x+\hat{b}_x^\dagger )$. 
(iii) The trap frequency $\nu_x$ and cavity field decay rate $\kappa$
are assumed to satisfy 
$\nu_x\gg\kappa\gg |(g_0\eta_x/\Delta ){\cal E}_{\rm L}(t)|$,
where $g_0$ is the single-photon atom-cavity mode coupling strength,
and $\Delta =\omega_{\rm a}-\omega_{\rm L}$.
The first inequality allows a rotating-wave approximation to be
made with respect to the trap oscillation frequency, while the
second inequality enables an adiabatic elimination of the cavity
field mode.

Under these conditions, the motional mode dynamics in the $x$ direction 
is well described by the simple quantum Langevin equation
\cite{Parkins99a}
\begin{equation} \label{QLE}
\dot{\tilde{b}}_x \simeq 
- \Gamma (t) \tilde{b}_x + \sqrt{2\Gamma (t)} \;
\tilde{a}_{\rm in}(t)\; ,
\end{equation}
where $\tilde{b}_x=e^{i\nu_xt}\hat{b}_x$ and 
$\Gamma (t)=[g_0\eta_x|{\cal E}_{\rm L}(t)|/\Delta ]^2/\kappa$.
The operator 
$\tilde{a}_{\rm in}(t)$ obeys the commutation relation 
$[\tilde{a}_{\rm in}(t),\tilde{a}_{\rm in}^\dagger (t^\prime )]
=\delta (t-t^\prime )$ and
describes the 
quantum noise {\em input to the cavity field}
(in a frame rotating at the cavity frequency).
From the linear nature of (\ref{QLE}), it follows
that the statistics of a (continuous) light field incident upon 
the cavity can be `written onto' the state of the atomic motion.
This also means that entanglement between separate light fields can be 
transferred to entanglement between separate motional states, as
we discuss below.
From a consideration of the input-output theory of optical
cavities \cite{Walls94}, it also follows that measurements on the 
cavity output field amount to measurements on the motion of the atom. 
In particular, one can show that
\begin{equation} \label{io}
\tilde{a}_{\rm out}(t) \simeq 
-\tilde{a}_{\rm in}(t) + \sqrt{2\Gamma (t)}\,
\tilde{b}_x(t) \, .
\end{equation}
So, for a vacuum input field, homodyne measurements 
on the cavity output field realize position or momentum 
measurements 
(or some mixture, depending on the local oscillator phase)
on the trapped atom. This enables the necessary
Bell-state analysis to be performed.

To begin the teleportation procedure,
Victor's atom is prepared in a particular motional
state $|\phi\rangle_{{\cal V}x}$ in the $x$ dimension
(e.g., by the techniques of \cite{Wineland98}). 
With the motion-light coupling switched off in Victor's cavity
(i.e., ${\cal E}_{{\rm L}{\cal V}}=0$),
this state is assumed to remain unchanged until required by
Alice for the Bell-state analysis.

Next, a position-momentum EPR state of Alice and 
Bob's atoms is prepared by using the motion-light
coupling described in (\ref{QLE}), with input light fields
from a nondegenerate optical parametric amplifier (NOPA).
This preparation, described in detail in \cite{Parkins99b}, is
depicted in Fig.~2(b).
The two quantum-correlated output light fields from a NOPA 
(operating below threshold) are separated and made to 
impinge on Alice and Bob's cavities, respectively.
Assuming $\Gamma_{\cal A}=\Gamma_{\cal B}=\Gamma$, after a
time $t\gg\Gamma^{-1}$, the following {\em pure} entangled motional 
state is prepared,
\begin{eqnarray} \label{psiAB}
|\psi\rangle_{\cal AB} &=& S_{\cal AB}(r) |0\rangle_{{\cal A}x} 
|0\rangle_{{\cal B}x} 
\nonumber
\\
&=& \left[ \cosh (r) \right]^{-1} \sum_{m=0}^\infty
\left[ -\tanh (r) \right]^m \, |m\rangle_{{\cal A}x} 
|m\rangle_{{\cal B}x} \, ,
\end{eqnarray}
where $|m\rangle_{{\cal A,B}x}$ are Fock states of the motional modes
and 
$S_{\cal AB}(r) = \exp [r(\tilde{b}_{{\cal A}x}\tilde{b}_{{\cal B}x} -
\tilde{b}_{{\cal A}x}^\dagger\tilde{b}_{{\cal B}x}^\dagger )]$,
with $r$ the `entanglement' parameter.
Once this state has been prepared, the atom-cavity couplings are
turned off ($\Gamma_{\cal A},\Gamma_{\cal B}\rightarrow 0$), as
is the NOPA pump field. 
Again, we assume that the entangled state (\ref{psiAB}) remains 
unchanged until the next step in the procedure.

At this stage in the protocol, the total system state is 
\begin{equation}
|\Psi_{\rm I}\rangle = |\phi\rangle_{{\cal V}x} \, 
|\psi\rangle_{\cal AB} \, .
\end{equation}
The Bell-state analysis performed by Alice is depicted in Fig.~3.
At a predetermined time, Victor switches on
his atom-cavity coupling $\Gamma _{{\cal V}}$ via 
${\cal E}_{{\rm L}{\cal V}}(t)$, thus converting the state 
$|\phi\rangle_{{\cal V}x}$ to that of a freely propagating field 
delivered to input beam-splitter BS of
Alice's sending station. With due accounting for propagation delay, Alice
has likewise switched on the coupling $\Gamma _{{\cal A}}$ from her
cavity, where, for simplicity, 
$\Gamma _{{\cal V}}=\Gamma _{{\cal A}}=\Gamma$.
Note that Victor and Alice's cavities both
have vacuum inputs at this stage. The two cavity output fields are combined 
by Alice at the 50/50 beamsplitter BS, the two outputs of which
are incident on homodyne detectors ${\rm D}_\pm$.
Through the input-output relation (\ref{io}), and through the mixing 
of the cavity output fields at the beamsplitter, these detectors 
effect homodyne measurements on the modes 
$\tilde{c}_\pm = 2^{-1/2} \,
( \tilde{b}_{{\cal V}x} \pm \tilde{b}_{{\cal A}x})$.
The effect of these measurements is to project the system
state onto quadrature eigenstates
of the modes $\tilde{c}_\pm$, given by
$|\chi_\pm\rangle_\pm = Q_\pm^\dagger (\chi_\pm )|0\rangle_\pm$, 
where
$Q_\pm^\dagger (\chi_\pm )=(2\pi )^{-1/4}\exp [
-(1/2)( \tilde{c}_\pm^\dagger e^{i\theta_\pm}
- \chi_\pm )^2+\chi_\pm^2/4]$,
with $\theta_\pm$ the local oscillator (LO) phases
\cite{Storey93,Herkommer96,Wiseman96}.
In \cite{Herkommer96,Wiseman96}, 
this projection is proved with the assumption
that the local oscillator photon flux matches the temporal shape
of the signal flux [which in our case is set by $\Gamma (t)$], 
while the variable $\chi$ is shown to be equivalent to the integrated 
homodyne photocurrent.

For the two homodyne measurements we choose LO
phases $\theta_+=0$ and $\theta_-=\pi /2$. With these choices
one can show that, in terms of the original mode operators,
\begin{eqnarray}
Q_+(\chi_+) && Q_-(\chi_-) = 
(2\pi )^{-1/2} \exp \left( -|\alpha |^2/2 \right) \nonumber
\\
&& \;\;\; \cdot \,
\exp \left( - \tilde{b}_{{\cal V}x}\tilde{b}_{{\cal A}x} 
+ \alpha \tilde{b}_{{\cal V}x}
+ \alpha^\ast \tilde{b}_{{\cal A}x} \right)  ,
\end{eqnarray}
where
$\alpha = (\chi_++i\chi_-)/\sqrt{2}$ .
The motional state of Bob's atom 
following the homodyne measurements, with results $\chi_\pm$,
can thus be written
\begin{eqnarray}
|\varphi\rangle_{{\cal B}x} &\propto & 
{}_{{\cal V}x}\langle 0| {}_{{\cal A}x}\langle 0| Q_+(\chi_+)
Q_-(\chi_-)|\Psi_{\rm I}\rangle \nonumber
\\
&\propto & 
{}_{{\cal V}x}\langle \alpha^\ast | {}_{{\cal A}x}\langle \alpha | 
\exp \left( - \tilde{b}_{{\cal V}x}\tilde{b}_{{\cal A}x} \right) 
|\Psi_{\rm I}\rangle .
\end{eqnarray}
Using properties of the squeezing operator $S_{\cal AB}(r)$ 
\cite{Walls94}, one can further reduce this to
\begin{equation} \label{phiB}
|\varphi\rangle_{{\cal B}x} \propto
{}_{{\cal V}x}\langle \alpha^\ast | 
\exp \left[ \Lambda \left( 
\tilde{b}_{{\cal V}x} - \alpha^\ast \right) 
\tilde{b}_{{\cal B}x}^\dagger 
\right] 
|\phi\rangle_{{\cal V}x} |0\rangle_{{\cal B}x} \, ,
\end{equation}
where $\Lambda = \tanh (r)$.
Expanding $|\phi\rangle_{{\cal V}x}$
in terms of the coherent states, i.e.,
$|\phi\rangle_{{\cal V}x} = \pi^{-1} \int d^2\beta \; 
{}_{{\cal V}x}\langle \beta |\phi\rangle_{{\cal V}x} \, 
|\beta\rangle_{{\cal V}x}$,
the right-hand-side
of (\ref{phiB}) becomes
\begin{eqnarray} \label{overlap}
&& \frac{1}{\pi} \int d^2\beta \;
{}_{{\cal V}x}\langle \beta |\phi\rangle_{{\cal V}x} \,
\frac{{}_{{\cal V}x}\langle \alpha^\ast |\beta\rangle_{{\cal V}x}}
{{}_{{\cal V}x}\langle \Lambda\alpha^\ast 
|\Lambda\beta\rangle_{{\cal V}x}} 
\nonumber
\\
&& \;\;\;\;\;\;\;\;\;\;\;\;\;\;\;\;\;\; \cdot \;
D_{\cal B}(-\Lambda\alpha^\ast ) 
D_{\cal B}(\Lambda\beta ) \, |0\rangle_{{\cal B}x} \, ,
\end{eqnarray}
where
$D_{\cal B}(\beta ) = \exp (\beta \tilde{b}_{{\cal B}x}^\dagger
- \beta^\ast \tilde{b}_{{\cal B}x})$
is the coherent displacement operator for Bob's atom.
In the limit of strong squeezing and entanglement 
($\Lambda\rightarrow 1$), (\ref{overlap}) approaches
\begin{equation}
D_{\cal B}(-\alpha^\ast ) \;
\frac{1}{\pi} \int d^2\beta \;
{}_{{\cal V}x}\langle \beta |\phi\rangle_{{\cal V}x} \, 
|\beta\rangle_{{\cal B}x} \, .
\end{equation}
That is, $|\varphi\rangle_{{\cal B}x}$ approaches a state which, apart
from a coherent displacement by $-\alpha^\ast$, is identical
to the initial motional state (in the $x$ dimension) 
of Victor's atom.

Given the measurement results $\chi_\pm$, transmitted
to Bob via a classical channel, the final step in the teleportation 
procedure is for Bob to apply a coherent displacement 
$\alpha^\ast$ (assuming $\Lambda\simeq 1$)
to the motional state of his atom, i.e.,
$D_{\cal B}(\alpha^\ast ) |\varphi\rangle_{{\cal B}x}\rightarrow
|\phi\rangle_{{\cal B}x}$.
In practice, this might be achieved by applying an electric
field (in the case of a trapped ion) along the $x$-axis
which oscillates at the trap frequency $\nu_x$, or, alternatively,
by applying off-resonant laser fields 
which drive stimulated Raman transitions between neighboring
trap levels \cite{Wineland98}.
After this, control of Bob's atom can be passed to Victor,
who is free to confirm the overall quality of the teleportation protocol, 
e.g., along the lines analyzed in \cite{Braunstein99}.

Issues of practicality associated with the motion-light state
transfer procedure central to our teleportation scheme 
have been discussed elsewhere \cite{Parkins99a,Parkins99b}.
In brief, desired conditions are of (i) strong coupling optical
cavity QED, such that $g_0^2/(\kappa\gamma )\gg 1$, where $\gamma$ 
is the atomic spontaneous decay rate,
and (ii) strong confinement of the atoms with minimal motional-state
decoherence.
Both of these conditions have been achieved separately
\cite{Monroe95,Wineland98,Hood98,Munstermann99}, 
and we expect that future experiments 
trapping single atoms inside optical cavities will be
able to meet these criteria simultaneously. 
Note that timescales for motional-state decoherence of trapped 
ions can be of the order of milliseconds \cite{Wineland98};
the typical timescale involved in our teleportation scheme,
$\Gamma^{-1}$, would likely be of the order of microseconds
\cite{Parkins99a,Parkins99b}. Finally, calculations in 
\cite{Braunstein98a} suggest reasonable (nonclassical) 
teleportation fidelities 
to be possible with values of the squeezing parameter $r>1$.

To conclude, we note that the scheme given here is just one of a
number of possibilities that we have analyzed. One could, e.g., 
eliminate Victor's atom and cavity and, as the state to be 
teleported, choose the motional state of Alice's atom along an axis 
{\em orthogonal} to the $x$-axis. 
After preparing the entangled ($x$-dimension) motional state 
$|\psi\rangle_{\cal AB}$ of Alice and Bob's atoms,
the orthogonal motional modes of Alice's atom could be linearly mixed
within the trap itself, 
in the fashion of a beamsplitter, using suitable
interactions with auxiliary laser fields 
\cite{Steinbach97,Parkins99c}, after 
which coupling to the cavity field and homodyne measurement of the
output light field would again provide the Bell-state analysis.
In addition, as we will discuss elsewhere \cite{Parkins99c},
it is possible to eliminate the NOPA from the scheme and 
use only trapped atoms interacting with cavity and 
laser fields both to produce and distribute the quantum entanglement 
required for the teleportation protocol.

ASP gratefully acknowledges support from the Marsden Fund of 
the Royal Society of New Zealand.
HJK is supported by the National Science Foundation, 
by DARPA via the QUIC Institute which is administered by ARO, 
and by the Office of Naval Research.

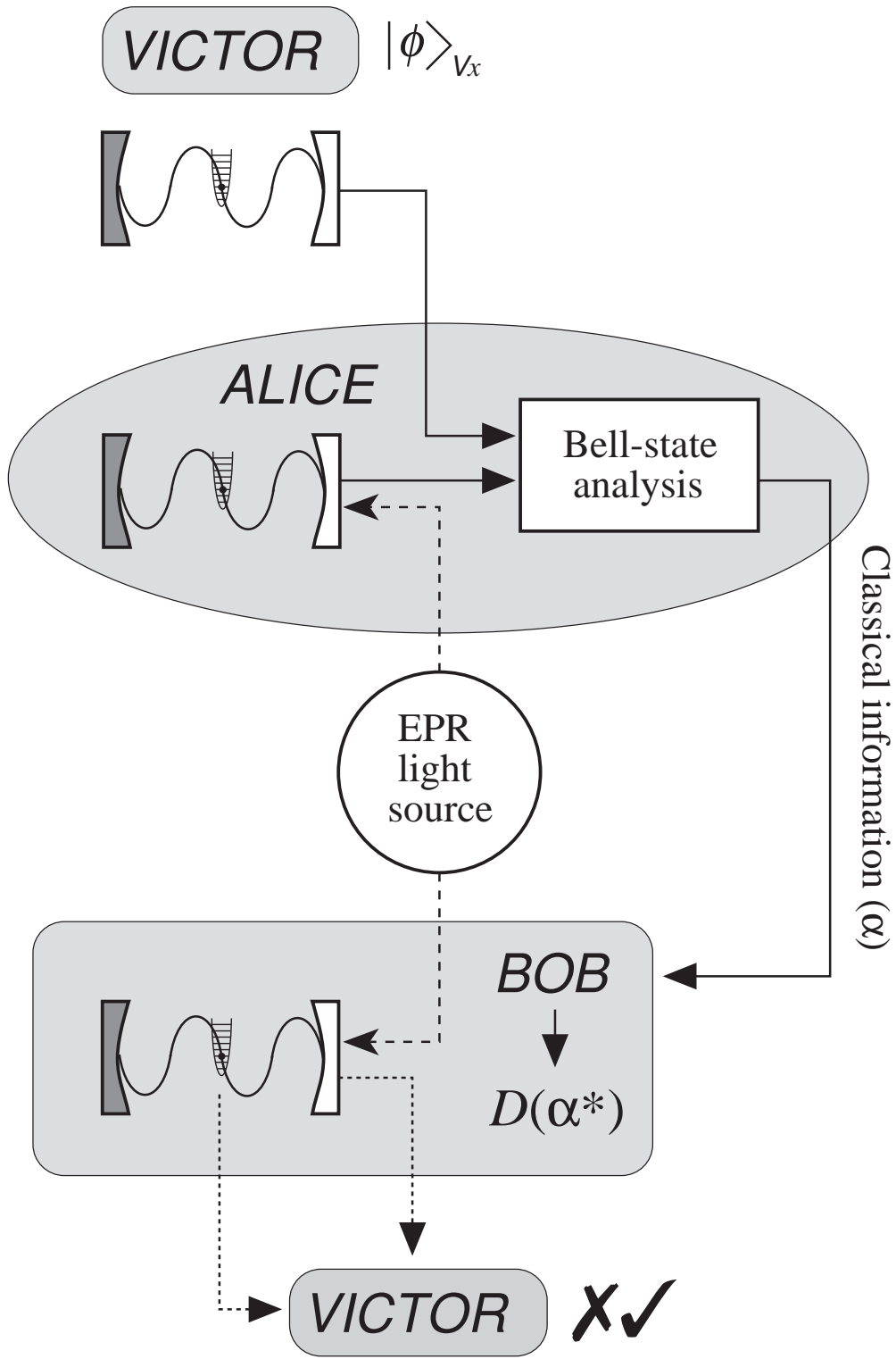
\begin{figure}
\caption{
Schematic of proposed teleportation scheme for atomic
wavepackets. 
Preparation of motional-state entanglement between Alice and Bob
and Bell-state analysis by Alice are facilitated by cavity-mediated
motion-light couplings.
}
\end{figure}

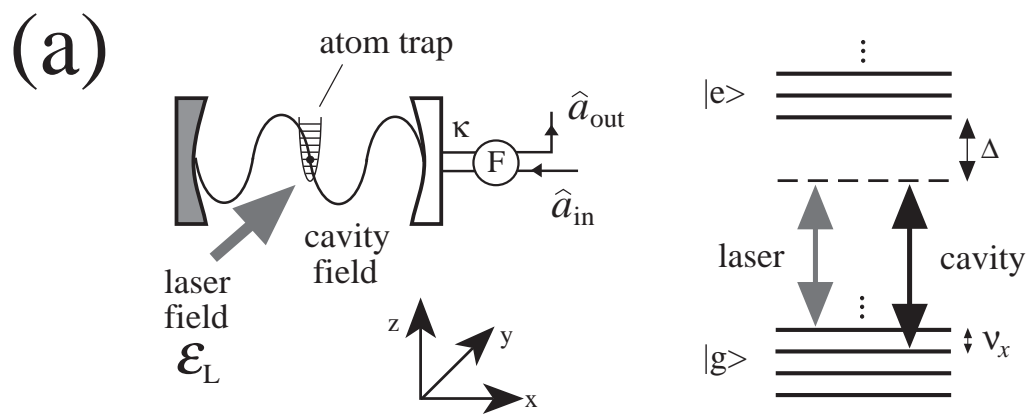
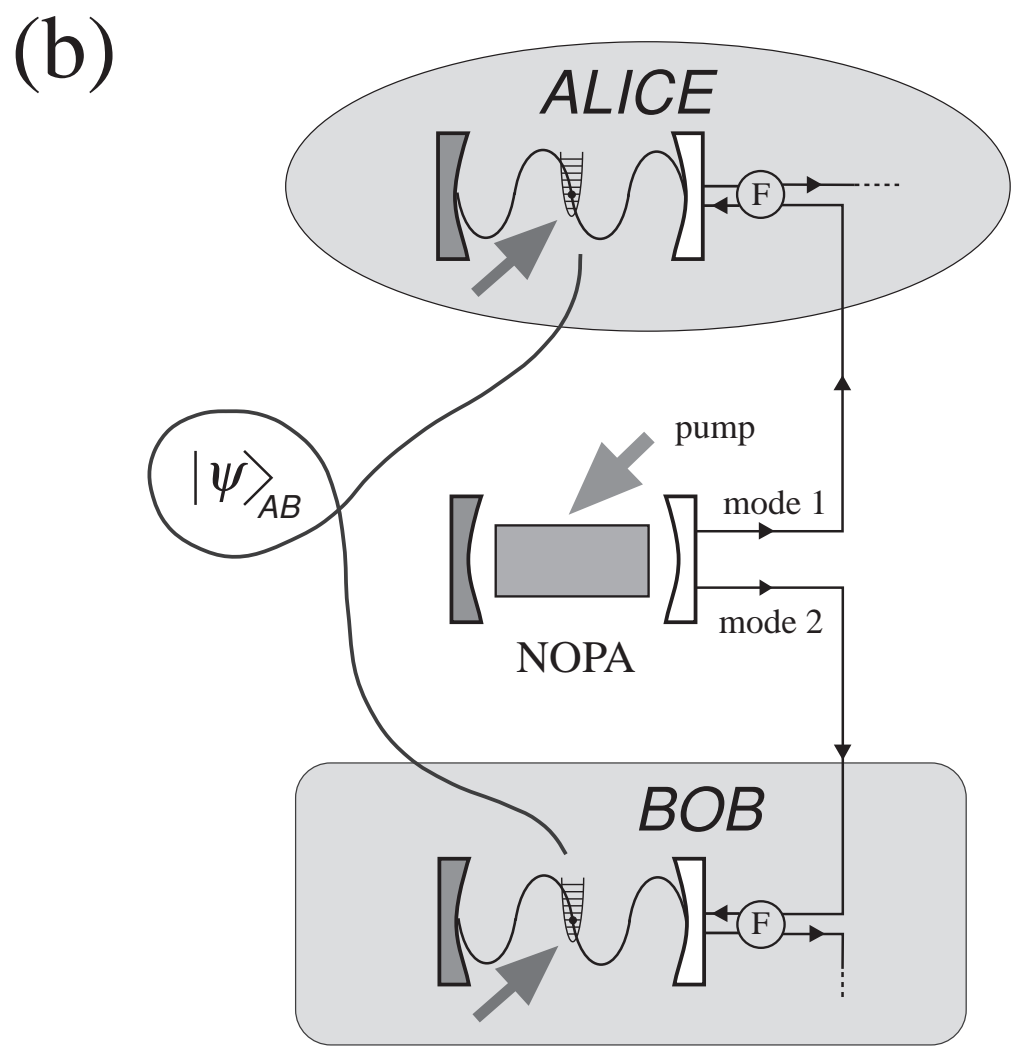
\begin{figure}
\caption{
(a) Proposed setup and excitation scheme 
for coupling between the motion of a trapped atom and a quantized
optical cavity mode, and thence to a freely propagating external
field. The cavity is 
assumed to be {\em one-sided}, i.e., one mirror is taken to be 
perfectly reflecting.
(b)
Preparation of a position-momentum EPR state of Alice and Bob's atoms. 
The two output fields from a nondegenerate parametric 
amplifier (NOPA) impinge on Alice and Bob's cavities,
respectively.
Faraday isolators (F) facilitate a unidirectional coupling between 
the entangled light source and the atom-cavity systems.
}
\end{figure}

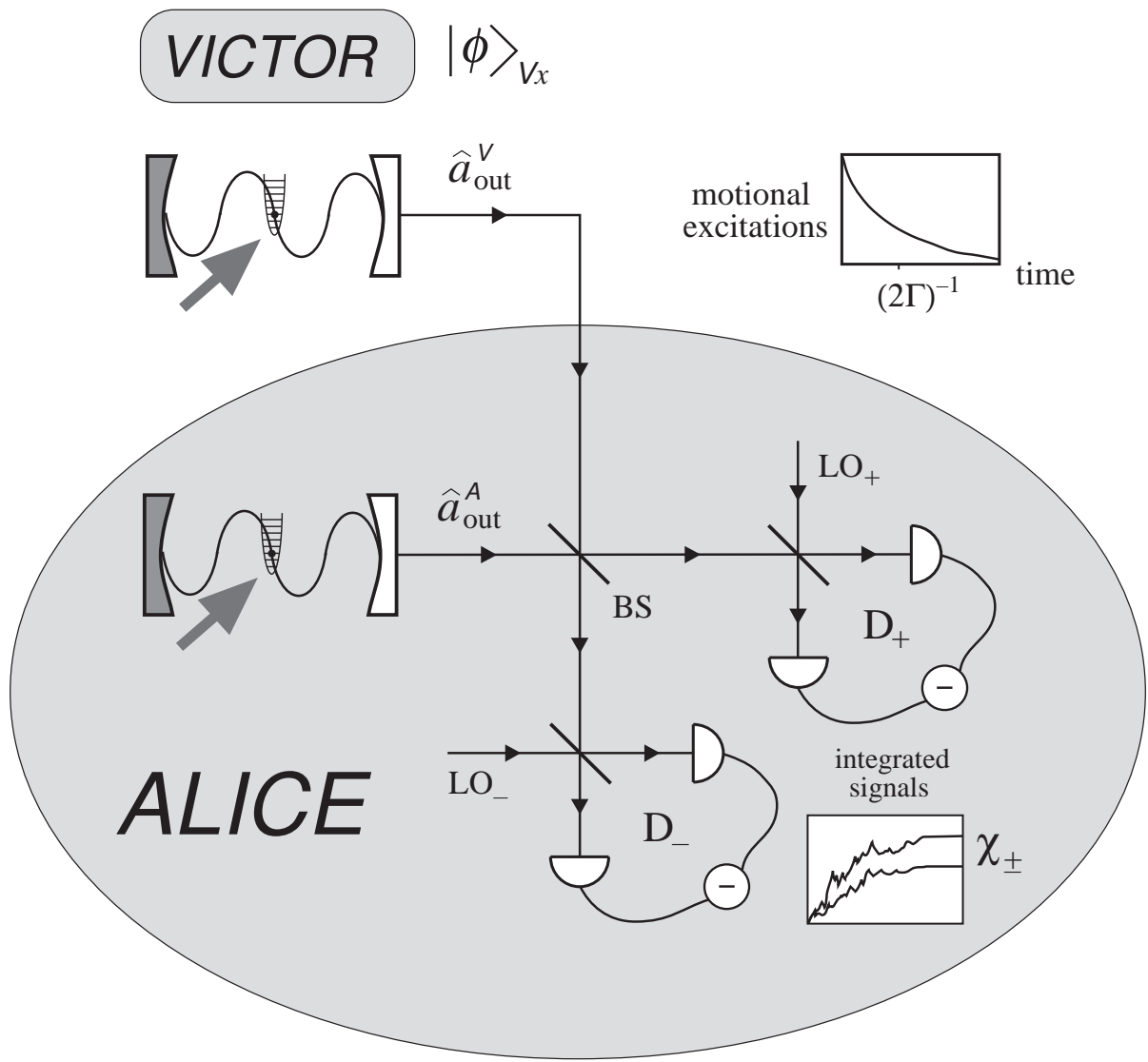
\begin{figure}
\caption{
Schematic of Alice's Bell-state analysis.
The output field representing Victor's unknown state is combined 
by Alice at a 50/50 beamsplitter (BS) with the output field from
her cavity. The resulting output fields from the BS are incident on
homodyne detectors ${\rm D}_\pm$.
The cavity output fields 
follow the motional modes, which decay on a timescale $\Gamma^{-1}$.
The local oscillator fields (${\rm LO}_\pm$) are pulsed, with 
temporal profiles chosen to match that of the cavity output fields.
}
\end{figure}

\end{document}